\documentclass[trackchanges, preprint]{aastex701}

\usepackage{amsmath}
\usepackage{bm}
\usepackage{xcolor}
\usepackage{CJK}

\begin{document}
\begin{CJK*}{UTF8}{gbsn}

\title{The Damping and Instability of Ion-acoustic Waves in the Solar Wind: Solar Orbiter Observations}

\author[orcid=0000-0002-8234-6480]{Hao Ran (冉豪)}
\affiliation{Mullard Space Science Laboratory, University College London, Dorking, RH5 6NT, UK}
\email[show]{hao.ran.24@ucl.ac.uk}

\author[orcid=0000-0002-0497-1096]{Daniel Verscharen}
\affiliation{Mullard Space Science Laboratory, University College London, Dorking, RH5 6NT, UK}
\email{d.verscharen@ucl.ac.uk}  

\author[orcid=0000-0002-2576-0992]{Jesse Coburn}
\affiliation{Blackett Laboratory, Imperial College London, London, SW7 2BW, UK}
\email{j.coburn@imperial.ac.uk}

\author[orcid=0000-0003-3623-4928]{Georgios Nicolaou}
\affiliation{Mullard Space Science Laboratory, University College London, Dorking, RH5 6NT, UK}
\email{g.nicolaou@ucl.ac.uk}

\author[orcid=0000-0003-4398-9931]{Charalambos Ioannou}
\affiliation{Mullard Space Science Laboratory, University College London, Dorking, RH5 6NT, UK}
\email{charalambos.ioannou.22@ucl.ac.uk}  

\author[orcid=0000-0001-7019-5905]{Xiangyu Wu (吴翔宇)}
\affiliation{Mullard Space Science Laboratory, University College London, Dorking, RH5 6NT, UK}
\email{xiangyu.wu.23@ucl.ac.uk}

\author[orcid=0009-0003-9856-5949]{Jingting Liu (刘婧婷)}
\affiliation{Mullard Space Science Laboratory, University College London, Dorking, RH5 6NT, UK}
\email{jingting.liu.22@ucl.ac.uk}

\author[orcid=0000-0001-6038-1923]{Kristopher Klein}
\affiliation{Lunar and Planetary Laboratory, University of Arizona, Tucson, AZ 85721, USA}
\email{kgklein@arizona.edu}

\author[orcid=0000-0002-5982-4667]{Christopher Owen}
\affiliation{Mullard Space Science Laboratory, University College London, Dorking, RH5 6NT, UK}
\email{c.owen@ucl.ac.uk}

\begin{abstract}

Observations of solar wind velocity distribution functions (VDFs) commonly reveal fine-scale structures.
These features strongly influence kinetic processes such as wave damping and instability, yet their role remains poorly understood.
We use a Gaussian Mixture Model (GMM) to separate proton and $\alpha$-particle (fully ionized helium) VDFs from Solar Orbiter Proton and Alpha-particle Sensor (PAS) measurements, and assess how measured VDFs affect the damping of compressive fluctuations with the Arbitrary Linear Plasma Solver (ALPS).
We analyze the dispersion relation and polarization properties of ion-acoustic (IA) waves in the solar wind.
Protons and $\alpha$-particles are represented by the measured VDFs derived from PAS observations.
For comparison, we also perform calculations using the bi-Maxwellian assumption for the VDFs.
Fine-scale structures of the measured proton VDFs reduce the damping rate of IA waves, even when $T_e \simeq T_i$.
In some cases, we find that the measured VDFs drive the IA mode unstable, while the corresponding bi-Maxwellian representations predict strong damping.
These results demonstrate that resolving the fine-scale structures of VDFs is essential for accurately capturing the kinetic physics of the solar wind.

\end{abstract}



\section{Introduction} 

The solar wind is a weakly collisional plasma, and its particle velocity distribution functions (VDFs) therefore usually depart from Maxwellian equilibrium \citep[e.g.,][]{vasyliunas1968survey, feldman1975solar, marsch1982solarhelium, marsch1982solar, marsch2006kinetic, livadiotis2013understanding, verscharen2019multi}.
These departures determine both the damping and instability of plasma waves, thereby defining the dissipation and transfer of energy across scales in the solar wind.
The resulting particle scattering acts as an effective viscosity, regulating the large-scale thermodynamic evolution of the plasma \citep{kunz2011thermally, kunz2014firehose, rincon2015non, riquelme2015particle, riquelme2018pic, coburn2022measurement}.
Previous studies often represent solar wind VDFs with simplified models such as bi-Maxwellian or $\kappa$-distributions \citep[e.g., ][]{hellinger2006solar, vstverak2008electron, livadiotis2009beyond, pierrard2010kappa, verscharen2017kinetic}.
These models capture some parametric non-equilibrium features quantitatively, including temperature anisotropies or non-thermal tails, but inevitably smooth over fine-scale velocity-space structures present in the VDF and visible in the measurements.
These fine-scale structures play a crucial role in wave--particle interactions and the kinetic physics of the solar wind \citep[e.g., ][]{isenberg2001heating, marsch2003ion, walters2023effects, coburn2024regulation}.

Compressive fluctuations in the solar wind, while of lower amplitude than Alfv\'enic fluctuations, modulate density and pressure and influence the plasma's thermodynamic evolution \citep{tu1995mhd, chen2016recent, verscharen2016collisionless, zhu2023regulation, ioannou2025polytropic}.
In magnetohydrodynamic (MHD) theory, the two propagating compressive modes are the slow and fast magnetosonic waves: density and magnetic-field fluctuations ($\delta n$ and $\delta |\bm{B}|$) are positively correlated for the fast mode and negatively correlated for the slow mode.
In-situ observations in the solar wind typically show anti-correlation between $\delta n$ and $\delta |\bm{B}|$, consistent with slow-mode-like behavior of the compressive fluctuations \citep{bavassano1989evidence, howes2012slow, klein2012using, yao2013small, verscharen2017kinetic}.
However, the collisionless nature of the solar wind requires a kinetic description.
In kinetic theory, the counterparts of the MHD slow magnetosonic mode are the ion-acoustic (IA) mode and the non-propagating mode \citep{verscharen2017kinetic}.
Previous studies based on bi-Maxwellian models for the VDFs suggest that IA waves experience strong Landau and transit-time damping unless $T_e \gg T_i$, which leads to the expectation that these waves should be absent when $T_e \simeq T_i$ \citep{barnes1966collisionless, feldman1975solar, gary1993theory}.
Nevertheless, IA-like compressive fluctuations are commonly observed in the solar wind, raising the question as to what enables their existence when $T_e \simeq T_i$, which is typical for the solar wind \citep[e.g., ][]{issautier2005solar, wilson2018statistical, salem2023precision}.

Numerical solvers for linear Vlasov--Maxwell theory  evaluate the kinetic wave dispersion relation, but most of them, including NHDS or PLUME \citep{verscharen2018nhds, klein2025plume}, rely on parametric representations of the VDFs, most commonly bi-Maxwellians for all involved plasma species.
Extensions to $\kappa$-distributions are implemented in solvers such as DSHARK \citep{astfalk2015dshark}.
In this work, we employ the Arbitrary Linear Plasma Solver \citep[ALPS; ][]{verscharen2018alps, klein_dielectric_2025}, which accepts arbitrary gyrotropic VDFs as the background distributions for the particle species.
This capability allows us to incorporate the measured VDFs into our linear Vlasov--Maxwell analysis, thereby accounting for their detailed velocity-space structure in determining the damping and growth rates of IA modes.
We evaluate the dispersion relation of IA waves using measured ion VDFs and compare the resulting predictions with direct observations of magnetic-field fluctuations. 
We identify the specific velocity-space structures in the measured VDFs that determine whether the mode is stable or unstable, thereby revealing how measured distributions regulate the behavior of compressive fluctuations in the solar wind.

In Section \ref{sec: data}, we present the data processing and VDF construction procedures, the numerical tools used to calculate the linear plasma response, and the quasi-linear framework for the analysis of wave growth.
The plasma behavior of IA modes associated with these VDFs is analyzed in Section \ref{sec:result}. 
We discuss and interpret our findings in Section~\ref{sec: discussion}, and summarize the main results in Section~\ref{sec: conclusions}.

\section{Data and Methodology}
\label{sec: data}

Observations for this study are obtained from the Proton and Alpha-particle Sensor (PAS) of the Solar Wind Analyzer \citep[SWA;][]{owen2020solar} suite, the  magnetometer \citep[MAG;][]{horbury2020solar}, and the Radio and Plasma Waves \citep[RPW;][]{maksimovic2020solar} instruments on board Solar Orbiter \citep{muller2020solar}. 
PAS is an electrostatic analyzer with a field of view spanning $-22.5^\circ$ to $22.5^\circ$ in elevation (9 bins) and $-24^\circ$ to $42^\circ$ in azimuth (11 bins).
It measures the three-dimensional VDF of protons and $\alpha$-particles within an energy range of 200~eV - 20~keV (96 bins) by sweeping through energy and elevation and recording the particle counts in each azimuth bin.
A full three-dimensional scan is completed in about 1~s, and we obtain one VDF every 4~s.
These counts maps are then converted into the VDF.
Magnetic-field vector measurements are taken from the level-2 MAG data set.
We use electron density estimates derived from RPW/BIAS probe-to-spacecraft potential measurements, which provide high time resolution and serve as a proxy for the plasma density during the wave analysis when the PAS cadence is insufficient \citep{khotyaintsev2021density}.
All data are available via the Solar Orbiter Archive\footnote{https://soar.esac.esa.int/soar/}.

\subsection{Separation of Protons and $\alpha$-particles}

\begin{figure}[!ht]
    \centering
    \includegraphics[width=1.0\linewidth]{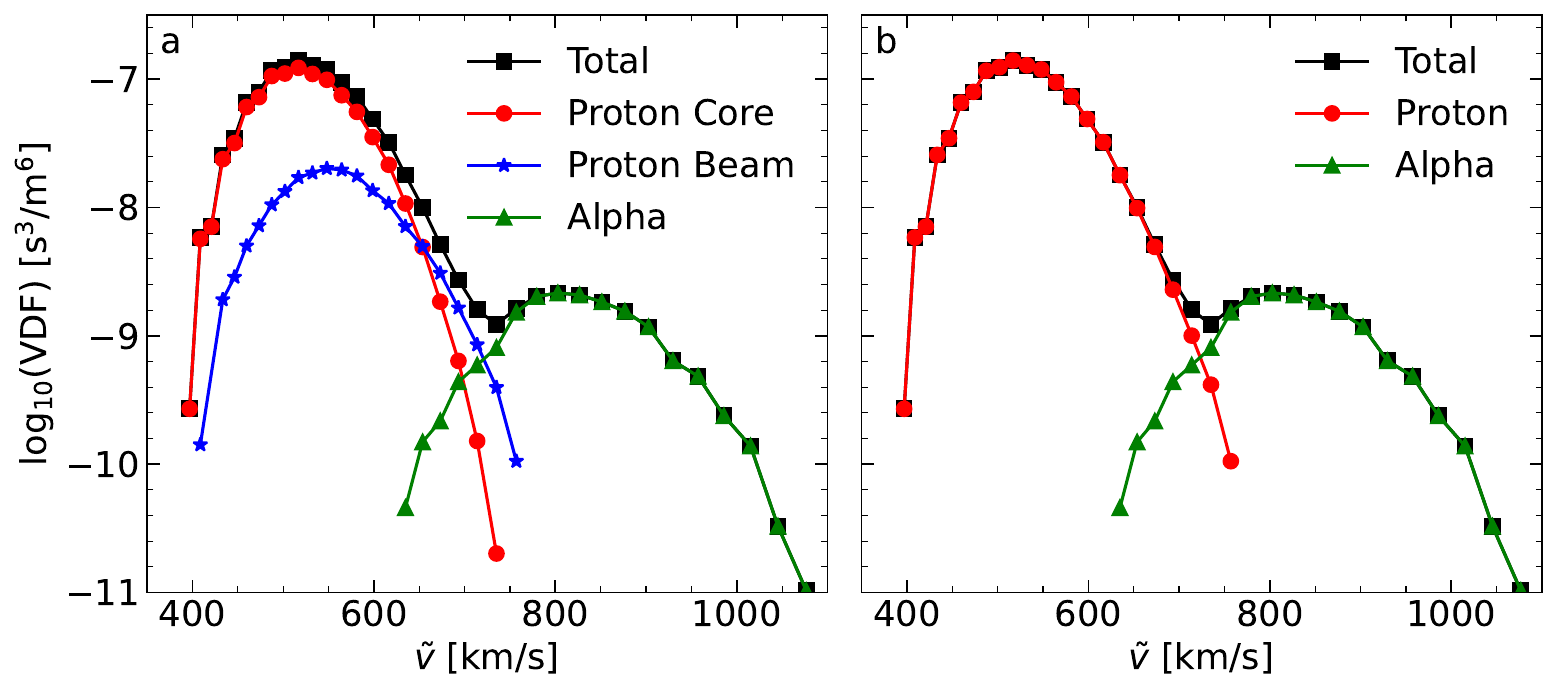}
    \caption{The averaged GMM separation of PAS measurements during the interval 02:06:00 - 02:07:00 UT on 2022 October 23.
    (a) VDF as a function ofthe proton-assumed pseudo-speed $\tilde{v}$, obtained by summing the measured VDF over elevation and azimuth. The total VDF (black squares) is decomposed into three GMM-identified components: proton core (red circles), proton beam (blue stars), and $\alpha$-particles (green triangles).
    (b) Same as (a) but combining the proton core and beam into a single proton population.}
    \label{fig:GMM_Separation}
\end{figure}

Typical PAS ion VDFs as a function of energy-per-charge exhibit several distinct peaks.
The strongest peak corresponds to the proton core, while a secondary proton beam is often present and drifts relative to the core along the direction of the local magnetic field at approximately the local Alfv\'en speed \citep{tu2004dependence, alterman2018comparison, verscharen2019multi}.
A separate $\alpha$-particle species appears at a higher energy-per-charge.
This displacement arises from two effects: first, $\alpha$-particles usually drift relative to the proton core along the direction of the magnetic field at roughly the local Alfv\'en speed  \citep{marsch1982solarhelium, kasper2006physics, ran2024alpha}; and second, their lower charge-to-mass ratio causes them to appear at twice the energy-per-charge of protons. 
Since velocity is derived from energy-per-charge assuming all particles were protons, the measured $\alpha$-particle velocities are overestimated by a factor of $\sqrt{2}$ compared to their true velocity.
We denote the proton-assumed measured pseudo-speed as $\tilde{v}$, such that $\tilde{v} = v$ for protons and $\tilde{v} = \sqrt{2} v$ for $\alpha$-particles, where $v$ is their respective true physical speed.

Because ALPS requires the VDFs of individual particle species, it is necessary to separate protons and $\alpha$-particles in PAS measurements.
In this work, we adopt the approach presented by \cite{de2023innovative}, who use a Gaussian Mixture Model \citep[GMM, ][]{reynolds2009gaussian} algorithm to separate the different species in the PAS measurements.

We first remove all points with just a single count from our VDFs, as these measurements have a 100\% relative uncertainty.
We also remove isolated pixels, defined as those with a measurement while all neighboring pixels are empty.
The input to the GMM algorithm is a five-dimensional dataset $\mathrm{X} = [\tilde{v}_\parallel, \tilde{v}_{\perp 1}, \tilde{v}_{\perp 2}, |\tilde{v}|, \mathrm{VDF}]$, where $\parallel$ and $\perp$ denote the directions parallel and perpendicular to the magnetic field.
All pseudo-velocities are expressed in the proton bulk frame defined using the ground-moment proton bulk velocity.
Since the GMM framework relies on linear covariance, it does not inherently recognize the nonlinear relationship between the Cartesian components and their Euclidean norm.
By explicitly providing $|\tilde{v}|$ as an input coordinate, we leverage the instrumental $\sqrt{2}$ scaling alongside the physical drifts to maximize the separation of the populations.
As input to our GMM model, we set the number of Gaussian components to $N=3$, representing the proton core, the proton beam, and the $\alpha$-particles.
Although our subsequent ALPS analysis utilizes only the combined total proton distribution, explicitly including the proton beam component is necessary for the clustering algorithm.
Without this explicit accounting for the proton beam component, the GMM algorithm often results in leakage of the identified proton cluster into the $\alpha$-particle component.
Initial values for the different components are assigned following the method of \cite{marsch1982solarhelium}.
Specifically, we first determine an energy-per-charge threshold $E_m$ by visually identifying the transition between the proton and $\alpha$-particle populations in the measured energy spectrum.
Measurements with energy-per-charge $E<E_m$ are assigned to protons ($f_{p, init}$), and those with $E>E_m$ to $\alpha$-particles ($f_{\alpha, init}$).

We then estimate the densities and the bulk velocities by integrating the initial VDFs:
\begin{equation}
    n_{j} = \int f_{j} (\bm{v}) d^3 v,
    \label{eq:0thmoment}
\end{equation}
\begin{equation}
    \boldsymbol{{U}}_{j} = \frac{1}{n_{j}} \int \bm{v} f_{j} (\bm{v}) d^3 v,
    \label{eq:1stmoment}
\end{equation}
where $f_j$ denotes the ion VDF of species $j$; in this step, $f_j$ corresponds to $f_{j, init}$.
For $\alpha$-particles, evaluating Equation~(\ref{eq:1stmoment}) within the proton-assumed velocity space yields their bulk pseudo-velocity $\boldsymbol{\tilde{U}}_{\alpha}$.
The initial mean vectors for the GMM components are set as: proton core: $\mathrm X=[0, 0, 0, 0, \max(f_{p, init}))]$, proton beam: $\mathrm X=[{V_{{A, p}}}, 0, 0, {V_{{A, p}}}, \max (f_{p, init}) / 10]$, and $\alpha$-particles: $\mathrm X=[({\tilde{U}_{\alpha, init \parallel}} - {U_{p, init \parallel}}), 0, 0, ({\tilde{U}_{\alpha, init \parallel}} - {U_{p, init \parallel}}), \max (f_{\alpha, init})]$, where ${V_{{A, p}}} = |\bm{B}| / \sqrt{4 \pi n_p m_p}$ is the local proton Alfv\'en speed calculated from the ground-moment proton density.
After the GMM separation, once the $\alpha$-particle parameters become available, the Alfv\'en speed is recomputed as ${V_{A}} = |\bm{B}| / \sqrt{4 \pi (n_p m_p + n_\alpha m_\alpha)}$ to account for their contribution and improve accuracy.
The initial VDF value for the proton beam is set to $\max(f_{p, init}) / 10$, reflecting its lower density compared to the core.

We implement the GMM using the \texttt{sklearn.mixture} module from the \texttt{scikit-learn} library \citep{pedregosa2011scikit}.
Once the model has converged, GMM returns for each pixel of the VDF the probabilities of belonging to the proton core, proton beam, or $\alpha$-particle components.
Multiplying these probabilities with the measured VDF values yields three separate three-dimensional VDFs, corresponding to the proton core ($f_{pc, g}$), proton beam ($f_{pb, g}$), and $\alpha$-particles ($f_{\alpha, g}$).

The model offers four covariance types: \textit{full} (each component has its own general covariance matrix), \textit{diag} (each component has its own diagonal covariance matrix), \textit{tied} (all components share the same general covariance matrix), and \textit{spherical} (each component has its own isotropic covariance matrix).
In our experience, the \textit{full} option usually fails to separate the components due to its complexity and limited data support.
Among the remaining options, we select the one that produces an $\alpha$-particle drift $(\boldsymbol U_{\alpha, g}-\boldsymbol U_{\mathrm p, g})$ most closely aligned with the magnetic field $\boldsymbol B$ \citep{marsch1982solarhelium, kasper2006physics, ran2024alpha}.

Finally, we extend this method to long PAS intervals using an iterative approach.
We begin by separating the ion components in the first time step of each time interval using the procedure described above.
The results from the initial particle separation are then used as the initial values for the next GMM iteration.
This process is repeated, with each time step using the previous particle separation as initial values, until all time steps of the interval have been processed.

Figure~\ref{fig:GMM_Separation} presents the averaged GMM separation result for the PAS interval under study.
In this example, taken from 02:06:00 to 02:07:00 UT on 2022 October 23, the 1-minute interval consists of 15 PAS measurements at 4~s cadence.
The method decomposes the measured ion VDF into three populations: the proton core, the proton beam, and the $\alpha$-particles.
The one-dimensional cuts show that each component occupies a well-defined region in velocity space.

\subsection{VDF Preparation for ALPS Analysis}

\begin{figure}[!ht]
    \centering
    \includegraphics[width=1.0\linewidth]{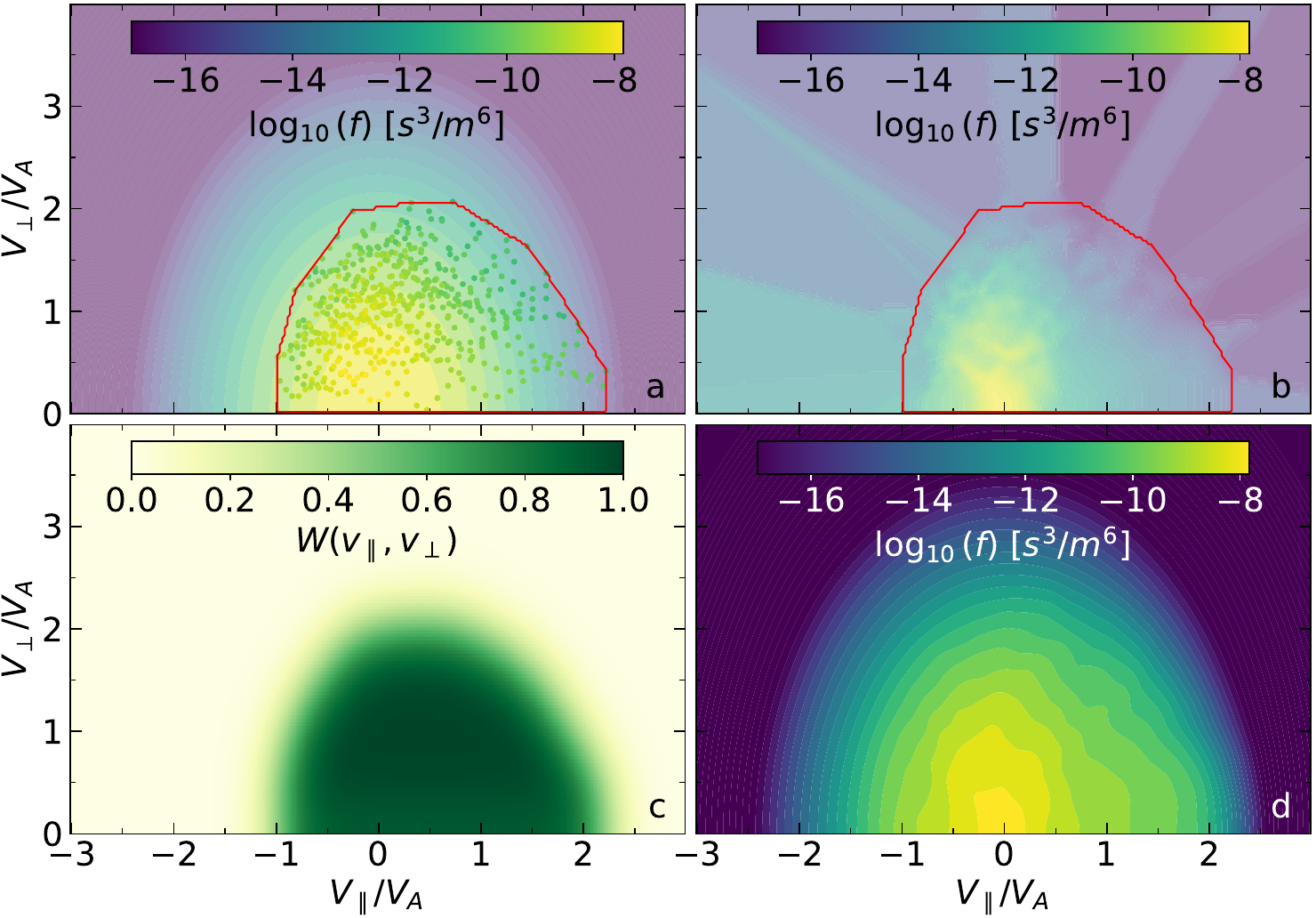}
    \caption{Construction of a collared VDF from the GMM-separated PAS proton measurements shown in Figure~\ref{fig:GMM_Separation}.
    (a) Measured VDF values are shown as scatter points, overlaid on a bi-Maxwellian background reconstructed from corresponding bulk moments according to Equation~(\ref{eq:bi_Max}). The red contour denotes the convex hull enclosing all measurements.
    (b) Linearly interpolated VDF based on the measured points.
    (c) Weight function $W(v_\parallel, v_\perp)$ generated by convolving the convex hull mask with a Gaussian kernel.
    (d) Final collared VDF obtained by blending the bi-Maxwellian background and the interpolated VDF using the weight function. 
    We use this shown VDF  as the input for our ALPS analysis.
    }
    \label{fig:VDF_Construction}
\end{figure}

PAS provides discrete VDF measurements, whereas ALPS requires a continuous VDF as input to solve the plasma dispersion relation.
To bridge this gap, we extend the measured velocity space beyond the PAS field of view by constructing a bi-Maxwellian ``collar'' around the measured region.

For this purpose, we first calculate the density $n_{j, g}$ and bulk velocity $\bm{{U}}_{j, g}$ from the GMM-decomposed VDFs $f_{j, g}$ using Equations~(\ref{eq:0thmoment}) and (\ref{eq:1stmoment}).
The parallel and perpendicular temperatures are then obtained as:
\begin{equation}
    T_{j,g \parallel} = \frac{m_j}{n_{j, g} k_B} \int c_\parallel^2 f_{j, g} d^3 v
    \end{equation}
    and
    \begin{equation}
    T_{j, g \perp} = \frac{m_j}{2 n_{j, g} k_B} \int c_\perp^2 f_{j, g} d^3 v,
    \label{eq:Tparperp}
\end{equation}
where $k_B$ is the Boltzmann constant, $\bm{c} = \bm{v} - \bm{{U}}_{j, g}$, $c_\parallel = \bm{c} \cdot \bm{\hat{b}}$, and $c_\perp^2 = c^2 - c_\parallel^2$.
Here $\bm{\hat{b}} = \bm{B} / |\bm{B}|$ denotes the unit vector along the magnetic field.
The corresponding thermal speeds are then given by:
\begin{equation}
    w_{j, g \parallel} = \sqrt{\frac{2 k_B T_{j, g \parallel}}{m_j}}, \quad 
    w_{j, g \perp} =  \sqrt{\frac{2 k_B T_{j, g \perp}}{m_j}}.
    \label{eq:thermalspeed}
\end{equation}

We then construct analytic bi-Maxwellian distributions for both protons and $\alpha$-particles in the form: 
\begin{equation}
    f_{j, bM} (\bm{v}) = \frac{n_{j, g}}{\pi^{3/2} w_{j, g \perp}^2 w_{j, g \parallel}} \exp \left( - \frac{v_\perp^2}{w_{j, g \perp}^2} - \frac{(v_\parallel - {U}_{j, g\parallel })^2}{w_{j, g \parallel}^2} \right).
    \label{eq:bi_Max}
\end{equation}
These serve as the background collars, approximating the expected VDF shape in regions of velocity space not directly sampled by PAS.
Panel (a) of Figure~\ref{fig:VDF_Construction} displays the GMM-separated proton measurement overlaid on the corresponding bi-Maxwellian collar.
The red curve outlines the convex hull enclosing all available data points. 
It represents the outer boundary of the measured velocity-space region.
It is computed using the \texttt{ConvexHull} function of the \texttt{scipy.spatial} Python library \citep{virtanen2020scipy}, which implements the Quickhull algorithm \citep{barber1996quickhull}.

Next, we perform a linear interpolation on all available measurements to fill velocity space between the PAS measurement points, as shown in panel (b) of Figure~\ref{fig:VDF_Construction}.
Due to the instrument's finite resolution, the measurement bins rarely align exactly with the $v_\perp = 0$ axis. 
Furthermore, because $v_\perp$ is bounded at 0, the singularity of the cylindrical coordinate system at $v_\perp=0$ causes standard interpolation algorithm to drop off and create unphysical empty regions near the axis.
To address this, we mirror the measurements into the $v_\perp < 0$ domain before interpolation and subsequently retain only the $v_\perp \geq 0$ region.
This procedure prevents artificial gaps and maintains physical consistency with the assumption of gyrotropy.


To ensure a smooth transition between the interpolated measured VDF ($f_{j, interp}$; panel (b) of Figure~\ref{fig:VDF_Construction}) and the bi-Maxwellian collar ($f_{j, bM}$; panel (a), we construct a soft blending weight based on the convex-hull mask.
The binary mask $M$, 1 for measured regions and 0 elsewhere, is first converted into a float array and then convolved with a broad Gaussian kernel:
\begin{equation}
    W (v_\parallel, v_\perp) = \left[ G_\sigma \star M \right] (v_\parallel, v_\perp),
    \label{eq:smooth_func}
\end{equation}
where $\star$ denotes convolution, and the Gaussian kernel is defined in pixel coordinates as:
\begin{equation}
     G_\sigma (i, k) = \frac{1}{2 \pi \sigma^2} \exp \left( - \frac{i^2 + k^2}{2 \sigma^2} \right),
     \label{eq:G_Kernal}
\end{equation}
$i$ and $k$ are pixel indices on the two-dimensional velocity grid of dimensions $240 \times 120$ (parallel $\times$ perpendicular), and $\sigma=10$ pixels specifies the kernel width.
We test the sensitivity of our results to $\sigma$ values ranging from 5 to 15 pixels and find that the results are insensitive to the specific choice of $\sigma$ in this range.
The convolution is implemented using the \texttt{scipy.ndimage.gaussian\_filter} \citep{virtanen2020scipy} routine.
This method results in a continuous weight field $W$ with values ranging from 1 (fully measured) to 0 (outside), as shown in panel (c) of Figure \ref{fig:VDF_Construction}.

The final $f_{j0}$ is obtained by:
\begin{equation}
   \log_{10} (f_{j0}) = W \log_{10} (f_{j, interp}) + (1 - W)  \log_{10} (f_{j, bM}),
\end{equation}
evaluated at each point in $(v_{\parallel},v_{\perp})$ space.
This blending avoids sharp discontinuities and ensures physical consistency, producing a smooth and stable hybrid distribution suitable for kinetic analysis.
Unlike the bi-Maxwellian approximation, this blended VDF passes through all measured points, providing a more realistic representation of the observed velocity distribution than parametric representations that are, for instance, based on moment fits to the measurements.
The final result in terms of $f_{j0}$ for this case is shown in panel (d) of Figure~\ref{fig:VDF_Construction}.

\subsection{Quasi-linear Framework and Growth-rate Calculation}

The slow, quasi-linear evolution of the background VDF $f_{0j}$  due to resonant wave--particle interactions is described by the quasi-linear diffusion equation \citep{stix1992waves}:
\begin{equation}
    \frac{\partial f_{0j}}{\partial t} = \frac{q_j^2}{8 \pi^2 m_j^2} \lim_{V \rightarrow\infty} \sum_{n=-\infty}^{\infty} \int \mathrm{d}^3 k \frac{1}{v_\perp V} \hat{G} \ v_\perp \delta (\omega_r - k_\parallel v_\parallel - n \Omega_j) |\psi^{j, n}| \hat{G} f_{0j},
    \label{eq:quasi_linear_f}
\end{equation}
where
\begin{equation}
    \hat{G} \equiv \left( 1 - \frac{k_\parallel v_\parallel}{\omega_r} \right) \frac{\partial}{\partial v_\perp} + \frac{k_\parallel v_\perp}{\omega_r} \frac{\partial }{\partial v_\parallel}, 
    \label{eq:G-operator}
\end{equation}

\begin{equation}
    \psi^{j, n} \equiv \frac{1}{\sqrt{2}} \left[ E_r e^{i \phi} J_{n+1} (\xi_j) + E_{l} e^{- i \phi} J_{n-1} (\xi_j) \right] + \frac{v_\parallel}{v_\perp} E_z J_n (\xi_j),
\end{equation}
$\Omega_j = (q_j B) / (m_j c)$ is the cyclotron frequency,  $\xi_j = k_\perp v_\perp / \Omega_j$ is the argument of the Bessel function $J_m$ of order $m$, and $n$ marks the resonance order.
The left- and right-hand components of the complex electric field amplitude $\vec E$ are defined by $E_l \equiv (E_x + i E_y)/\sqrt{2}$ and $E_r \equiv (E_x - i E_y)/\sqrt{2}$.
The $\delta$-function in Equation~(\ref{eq:quasi_linear_f}) ensures that only particles fulfilling the resonance condition
\begin{equation}
    \omega_r = k_\parallel v_\parallel + n \Omega_j
    \label{eq:resonance}
\end{equation}
contribute to the resonant wave--particle interactions at a given $n$. 
During Landau-resonant ($n=0$) interactions, particles with $v_\parallel = \omega_r / k_\parallel$ interact mostly with the parallel electric field $E_z$ of the wave.
In quasi-linear theory, this produces diffusion primarily along the magnetic field in velocity space.
In contrast, cyclotron-resonant ($n \neq 0$) interactions lead to diffusion paths that are tangent to curves of constant kinetic energy in the wave frame.
These trajectories are described by
\begin{equation}
    (v_\parallel - v_{ph})^2 + v_\perp^2 = \text{constant},
    \label{eq:conserve_Ek}
\end{equation}
where $v_{ph} \equiv \omega_r / k_\parallel$ is the field-parallel phase speed of the resonant waves.
This behavior reflects that quasi-linear diffusion conserves particle kinetic energy in a frame moving with velocity $v_{ph} \hat{\bm{b}}$.

Each particle species $j$ contributes to the total growth/damping rate with
\begin{align}
    \frac{\gamma_j}{|\omega_r|} = \frac{\pi}{8 n_j} \left| \frac{\omega_r}{k_\parallel} \right| \left( \frac{\omega_{pj}}{\omega_r} \right)^2 \sum_{n=-\infty}^{\infty} \int_{0}^\infty d v_\perp v_\perp^2 \int_{-\infty}^\infty d v_\parallel \ \delta \left( v_\parallel - \frac{\omega_r - n \Omega_p}{k_\parallel} \right) \frac{|\psi_n^2| \hat{G} f_{0j}}{\mathcal{E}},
    \label{eq:growth_rate}
\end{align}
where $\omega_{pj} = \sqrt{4 \pi n_j q_j^2 /m_j}$ is the plasma frequency of species $j$ and $\mathcal{E}$ is the wave energy density \citep[e.g.,][]{kennel1967resonant, stix1992waves}.
The total growth/damping rate of the mode is obtained by summing the individual contributions over all particle species:
\begin{equation}
    \gamma = \sum_j \gamma_j.
\end{equation}
This quasi-linear framework is valid in the weak-growth limit $|\gamma| \ll |\omega_r|$, and here we restrict our analysis to $|\gamma / \omega_r| < 1 / e$ \citep{walters2023effects}.
For a comprehensive discussion of this framework and its application to instabilities in the solar wind, we refer the reader to \cite{verscharen2022electron}.

\subsection{Calculation of the Dispersion Relation}
\label{sec: ALPS_method}

We use ALPS to analyze the linear kinetic behavior of the plasma.
ALPS finds solutions to the linear Vlasov--Maxwell dispersion relation by determining the real frequency $\omega_r$ and growth/damping rate $\gamma$ that satisfy the wave equation $\mathcal{D}=0$ for specific background distributions $f_{0j}$.
It employs an iterative Newton-secant method \citep{press1992art} to update $\omega_r$ and  $\gamma$ based on initial guesses for their values.
Each $f_{0j}$ is expressed in gyrotropic momentum space, where the particle momentum is defined as $\bm{p}_j = m_j \bm{v}$.
The distribution is provided in an input ASCII table (the $f_0$-table) listing the $\bm{p}_\parallel$, $\bm{p}_\perp$, and the corresponding values of $f_{0j}$. 
The table covers $0 \leq \bm{p}_\perp \leq \bm{p}_{\max, \perp j}$ and $- \bm{p}_{\max, \parallel j} \leq \bm{p}_{\parallel} \leq \bm{p}_{\max, \parallel j}$ over $N_\perp$ and $N_\parallel$ steps in each dimension.
The algorithm then uses a second-order finite differencing scheme to compute $\partial f_{0j} / \partial p_\perp$ and $\partial f_{0j} / \partial p_\parallel$.
The integrals in the susceptibility calculation (Eq.~(2.9) by \cite{verscharen2018alps}) are performed independently for each species $j$ and resonance order $n$, which enables ALPS to parallelize their calculation.

We set the numerical parameters following the definitions introduced by \cite{verscharen2018alps}.
The velocity grid resolution is $N_{\parallel}=240$ points in the parallel direction and $N_{\perp}=120$ points in the perpendicular direction.
We set the Bessel function truncation threshold to $J_{\max} = 10^{-45}$.
The initial values are chosen to satisfy the expected IA dispersion relation:
\begin{equation}
    \omega_r \simeq k_\parallel c_s
    \label{eq:IA_dispersion}
\end{equation}
where 
\begin{equation}
    c_s = \sqrt{\frac{ n_ek_BT_e + 5(n_p k_BT_p + n_\alpha k_BT_\alpha)/3}{n_p m_p + n_\alpha m_\alpha + n_e m_e}}
    \label{eq:sound_speed}
\end{equation}
is the ion sound speed.

We compare four reference cases using different combinations of ion $f_0$-tables derived from PAS:
(i) both protons and $\alpha$-particles are represented by their collared measured distributions, 
(ii) $\alpha$-particles are represented by their collared measured distribution and protons as bi-Maxwellians,
(iii) protons are represented by their collared measured distribution and $\alpha$-particles as bi-Maxwellians,
and (iv) both species are represented as bi-Maxwellians.
Here, the bi-Maxwellians are constructed from the corresponding moments of the PAS measurements according to Equation~(\ref{eq:bi_Max}).
In all cases, electrons are modeled as an isotropic Maxwellian with the same total temperature as the protons ($T_e = T_p$), and their densities and drift speeds are set to ensure quasi-neutrality and a zero net current along the background magnetic field in the plasma.

\subsection{Wavelet Coherence Analysis}
\label{sec: coherence_method}
To characterize the phase relationship and linear coupling between density and magnetic-field fluctuations associated with the identified mode, we compute the wavelet coherence between plasma density and the parallel magnetic field component.
We take the plasma density ($n_{rpw}$) as the electron density provided by RPW, while the magnetic field is measured by MAG.
We construct $B_{\parallel} (t)$ by projecting the magnetic field onto the local mean-field direction obtained with a centered rolling window of length 30~s.
We denote the continuous wavelet transforms of the density and parallel magnetic-field fluctuations at scale $s$ and time $t$ as $C_n(s, t)$ and $C_B(s, t)$.
The cross-wavelet spectrum is defined as $C_{nB} = C_n C^*_B$, where the asterisk denotes the complex conjugate.
The squared wavelet coherence is then \citep{torrence1998practical}:
\begin{equation}
    R^2 = \frac{|\langle s^{-1} C_{nB} \rangle|^2}{\langle s^{-1} |C_n|^2 \rangle \langle s^{-1} |C_B|^2 \rangle},
    \label{eq:coherence_R2}
\end{equation}
where $\langle \cdot \rangle$ indicates a smoothing operator in both $s$ and $t$. 
To determine the phase relationship between the two signals, we compute the cross-wavelet phase angle:
\begin{equation}
    \Delta\phi = \tan^{-1} \left( \frac{\Im(C_{nB})}{\Re(C_{nB})} \right).
    \label{eq:coherence_phase}
\end{equation}
The wavelet coherence analysis is carried out using the Python package \texttt{pycwt} \citep{pycwt}.
This method provides a time-frequency map of the linear correlation between $\delta n_{rpw}$ and $\delta B_\parallel$.

\section{Results}
\label{sec:result}

We first present our analysis of the wave dispersion relation and the resonance behavior, before inspecting the velocity-space dynamics of the resonant wave--particle interactions in more detail. We then compare the linear and quasi-linear Vlasov--Maxwell results with wave observations.

\subsection{Dispersion Relation and Resonance Analysis}
\label{sec: Dispersion_resonance}

\begin{figure}[!ht]
    \centering
    \includegraphics[width=1.0\linewidth]{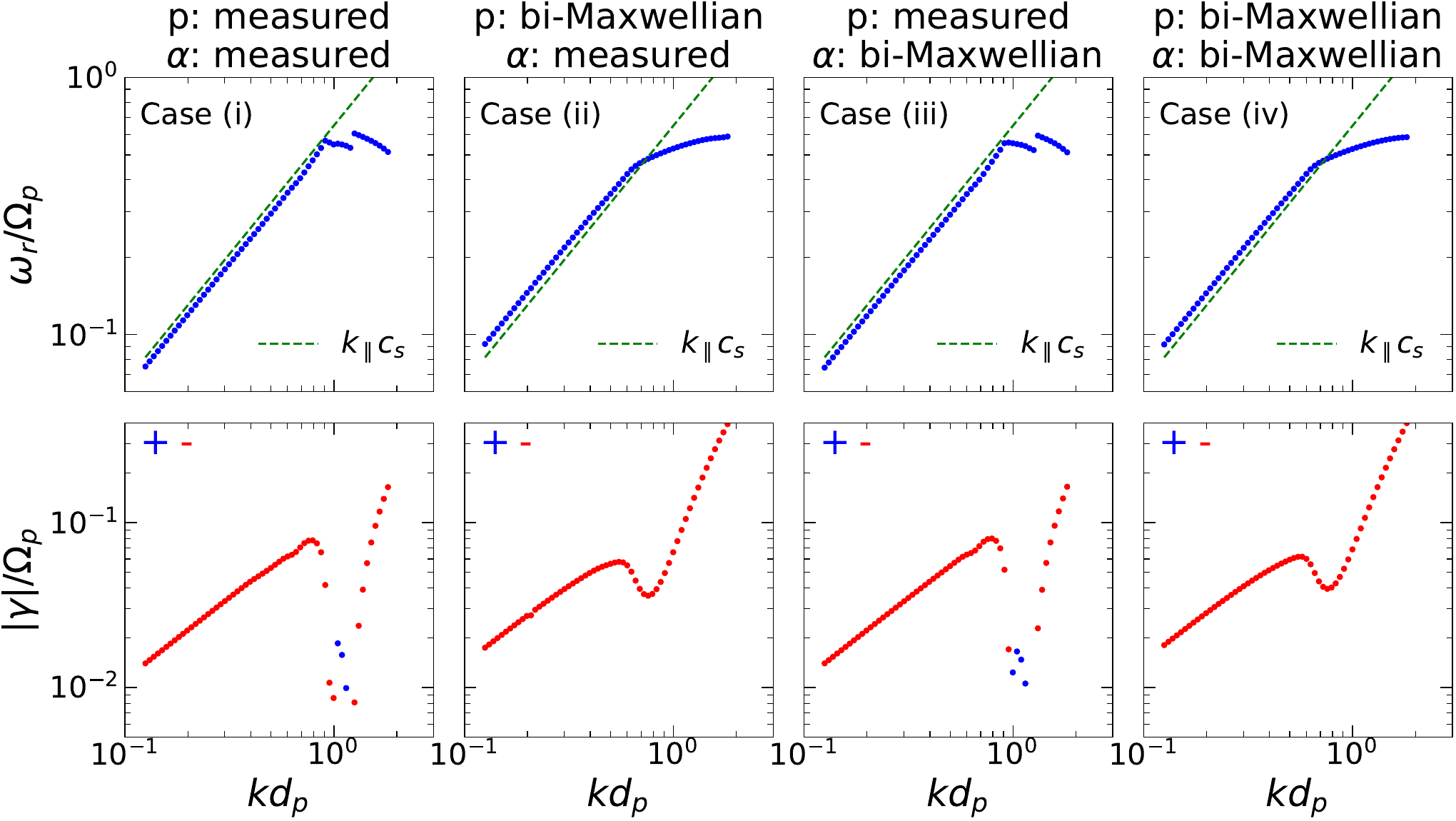}
    \caption{ALPS solutions for the real ($\omega_r$) and imaginary ($\gamma$) parts of the wave frequency during the interval 02:06:00 - 02:07:00 UT on 2022 October 23, under different assumptions for proton and $\alpha$-particle background VDFs.
    Columns (left to right): (i) measured VDFs for both species; (ii) bi-Maxwellian protons with measured $\alpha$-particles; (iii) measured protons with bi-Maxwellian $\alpha$-particles; (iv) bi-Maxwellian VDFs for both species.
    Top row: real frequency $\omega_r / \Omega_p$ vs. $k d_p$, with the green dashed line indicating the IA dispersion from Equation~(\ref{eq:IA_dispersion}).
    Bottom row: growth/damping rates $\gamma / \Omega_p$, with red indicating damping ($\gamma < 0$) and blue indicating instability ($\gamma > 0$).
    }
    \label{fig:dispersion_relation_221023}
\end{figure}

Figure~\ref{fig:dispersion_relation_221023} shows the ALPS dispersion solutions for the four reference cases defined in Section~\ref{sec: ALPS_method}, corresponding to cases (i)-(iv) in the figure.
Both the real frequency $\omega_r$ and the growth rate $\gamma$ are normalized to the proton cyclotron frequency $\Omega_p$ and shown as functions of the normalized wavenumber $k d_p$, where $d_p = {V_A} / \Omega_p$ denotes the proton inertial length.

During this interval, the GMM-separated proton bulk speed is 529.22~km/s, the number densities of protons and $\alpha$-particles are 31.86~cm$^{-3}$ and 0.83~cm$^{-3}$, respectively. We obtain $T_{\perp p} / T_{\parallel p}=2.06$, $T_{\perp \alpha} / T_{\parallel \alpha}=0.92$, $U_{\alpha\parallel}=90.96$~km/s, ${V_A} = 108.62$~km/s, and $c_s = 85.52$~km/s.
For all VDF inputs, ALPS scans over $k d_p$ from 0.1 to 5.0 at a fixed angle $\theta_{kB} = 30^\circ$ between $\boldsymbol k$ and $\boldsymbol B$.
This angle is chosen based on preliminary scans performed at steps of $10^\circ$ from quasi-parallel to quasi-perpendicular wave vectors, which show that $\theta_{kB}=30^\circ$ yields the maximum growth rate of the IA mode in this case.

\begin{figure}[!ht]
    \centering
    \includegraphics[width=1.0\linewidth]{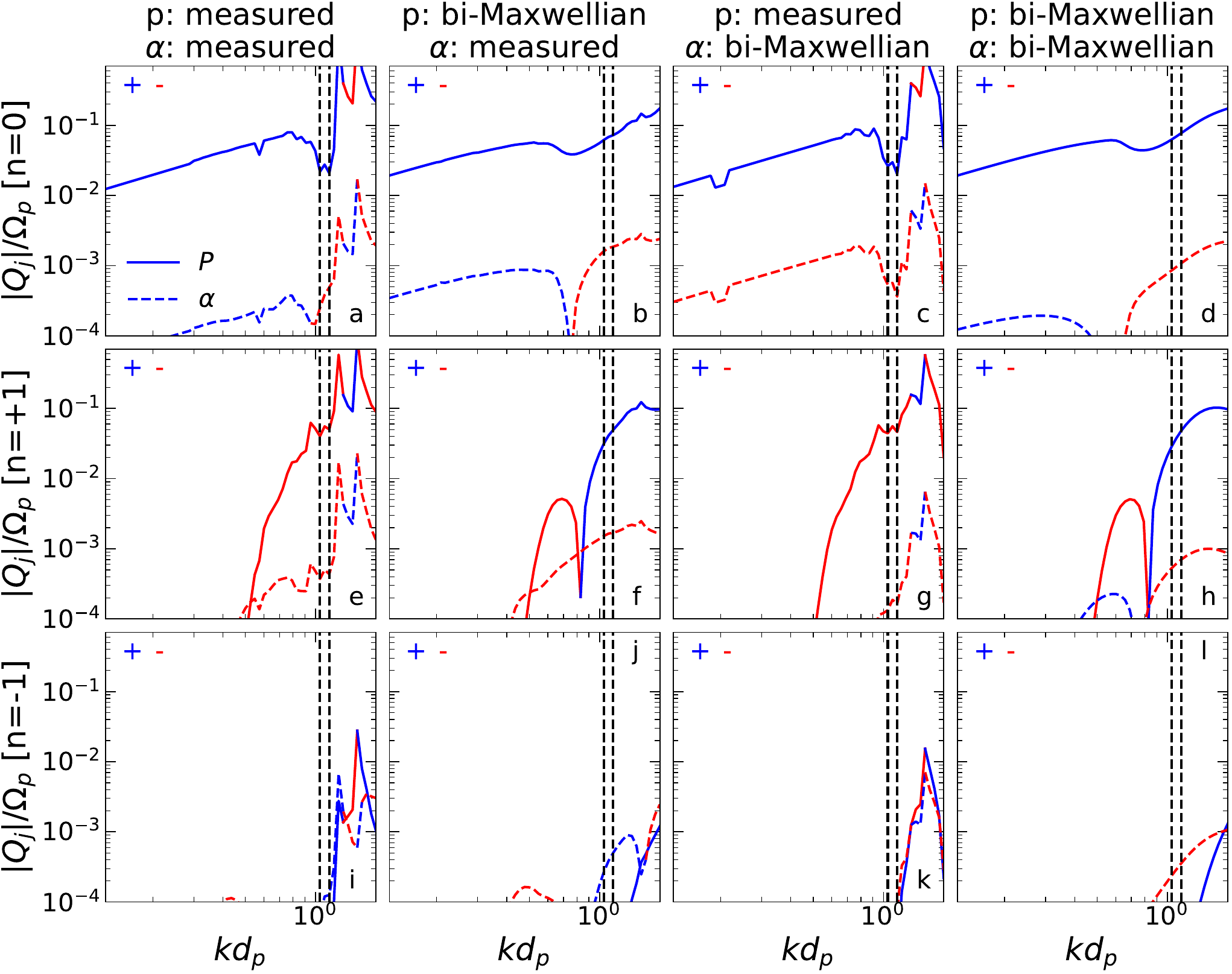}
    \caption{Heating rates of protons (solid curve) and $\alpha$-particles (dashed curve) through $n=0$ (top row), $n=+1$ (middle row), and $n=-1$ (bottom row) resonances.
    Blue indicates positive heating (wave energy transferred to particles; wave damping), while red indicates negative heating (particles driving the wave, instability).
    From left to right: same cases as Figure (\ref{fig:dispersion_relation_221023}).
    Vertical dashed lines mark the wavenumbers in which $\gamma > 0$ in case (i).
    }
    \label{fig:heating_rates}
\end{figure}

The identified mode corresponds to the IA branch in all configurations, as confirmed by its phase speed and the anti-correlation between density and magnetic-field fluctuations (correlation coefficient -0.698, not shown).
Across most wavenumbers, the growth rate is negative, but it becomes positive near $k d_p \simeq 1$ only when the measured proton VDF is used (cases (i) and (iii) in the first and third columns of Figure~\ref{fig:dispersion_relation_221023}).
In these cases, the dispersion relation and growth rate are different from those obtained with a bi-Maxwellian representation of protons (cases (ii) and (iv) in the second and fourth column), whereas representing $\alpha$-particles with a bi-Maxwellian produces little change, as seen by comparing cases (i) and (iii), as well as cases (ii) and (iv).

To investigate the origin of this instability, we examine the species-specific heating rates $Q_j$ from different resonance mechanisms in Figure~\ref{fig:heating_rates}.
The heating rates quantify the net energy exchange between each species and the wave through Landau ($n = 0$) and first-order cyclotron ($n = \pm 1$) resonances.
They are computed following the formulation given in Appendix~B of \cite{klein_dielectric_2025}.
A positive heating rate ($Q_j>0$) indicates that a species gains energy from the waves, leading to wave damping, whereas a negative heating rate ($Q_j<0$) corresponds to energy transfer from particles to waves, driving instability. 
The black vertical dashed lines indicate the $k d_p$-range in which $\gamma > 0$ in the fully measured configuration of case (i) as shown in the first column of Figure~\ref{fig:dispersion_relation_221023}.

For the $n=0$ resonance (top row), protons experience net heating across the entire wavenumber range, indicating that they predominantly act as a sink of wave energy and thus contribute to damping.
When measured proton VDFs are used (cases (i) and (iii) in Figure~\ref{fig:heating_rates}), the heating profile shows a pronounced dip where the total growth rate is positive (between the vertical dashed lines), representing a local reduction of proton heating at this wavenumber.
By contrast, the heating rates of $\alpha$-particles are consistently much smaller in magnitude than those of the protons.

As for $n=+1$ resonance (middle row), when protons are represented by a bi-Maxwellian (cases (ii) and (iv) in Figure \ref{fig:heating_rates}), they lose energy at $0.5 \lesssim k d_p \lesssim 0.8$, and gain energy at larger $k d_p$.
For the measured distributions (cases (i) and (iii) in Figure~\ref{fig:heating_rates}), proton cooling begins around $k d_p \sim 0.6$ and remains high over larger wavenumbers.
Within the $k d_p$-range between the vertical dashed lines, the proton heating rate is negative when using the measured distributions (cases (i) and (iii)), whereas it is positive when protons are represented by bi-Maxwellian (cases (ii) and (iv)).
The $n=-1$ resonance (bottom row) contributes only weakly, with proton and $\alpha$-particle heating rates both remaining at low and comparable levels.

\subsection{Velocity-space Dynamics}
\label{sec: velocity_space_diffusion}

\begin{figure}[!ht]
    \centering
    \includegraphics[width=1.0\linewidth]{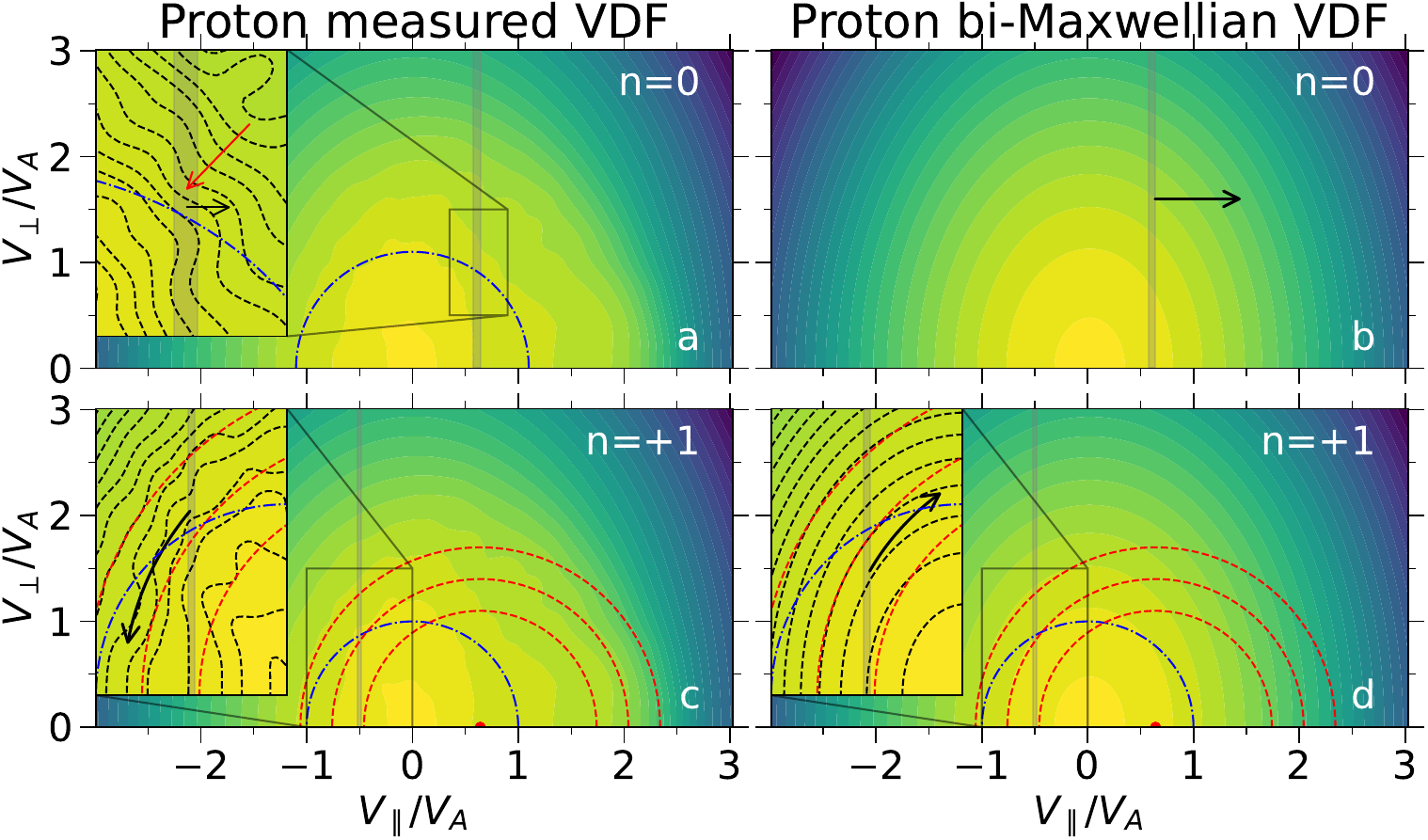}
    \caption{Diffusive flux of particles in velocity space under $n=0$ (top row) and $n=+1$ (bottom row) resonances, shown for measured (left) and bi-Maxwellian (right) VDFs.
    The red arrow in panel (a) marks the regions in which the VDF gradient is softened.
    Gray vertical bands indicate the resonance speeds corresponding to wavenumbers where $\gamma > 0$ in case (i).
    Blue dash-dotted semicircles denote surfaces of constant kinetic energy.
    Red dashed semicircles represent Equation~\ref{eq:conserve_Ek}, and black arrows illustrate the expected quasi-linear diffusion paths in velocity space.
    }
    \label{fig:resonance_path_221023}
\end{figure}

We evaluate the resonance condition in Equation~(\ref{eq:resonance}) to identify where in velocity space resonant interactions occur.
Figure~\ref{fig:resonance_path_221023} overlays this information onto two-dimensional representations of the bi-Maxwellian and measured proton VDFs.
The blue dash-dotted semi-circles denote surfaces of constant kinetic energy (i.e., $v_{\perp}^2+v_{\parallel}^2=\text{constant}$).
When resonant particles move inward across a blue semi-circle, their kinetic energy decreases, and energy is transferred to the waves, driving instability.
Conversely, outward motion corresponds to wave damping.
The red dashed semi-circles trace diffusion paths  as defined by Equation~(\ref{eq:conserve_Ek}) associated with cyclotron resonance.
The direction of the diffusive particle transport through velocity space along these paths (i.e., clockwise vs. counter-clockwise) is determined by the gradient of the VDF through $\hat{G}$ in Equation~(\ref{eq:quasi_linear_f}). The particles must diffuse from regions of higher to lower VDF values \citep{verscharen2013dispersion, verscharen2019multi, verscharen2022electron}.
The corresponding directions are indicated by the black arrows.
The gray bands indicate the resonance speeds $v_{res} = (\omega_r - n \Omega_p) / k_\parallel$ associated with the wavenumbers for which the mode is unstable ($\gamma > 0$) in case (i) of Figure~\ref{fig:dispersion_relation_221023}.
In panels (a) and (b), the resonance corresponds to the $n=0$ resonance, such that the resonance speed reduces to the phase speed, $v_{res} = v_{ph} = \omega_r / k_\parallel$.

Panels (a) and (b) show the Landau resonance with $n=0$.
In the measured VDF (panel (a)), the red arrow highlights a region in which the velocity gradient is softened within the resonant velocity range (gray span). 
In contrast, this structural feature is absent in the bi-Maxwellian case (panel (b)), which exhibits a steep gradient across the same $v_\parallel$ interval.
Panels (c) and (d) show the $n=+1$ cyclotron resonance.
The measured VDF in panel (c) displays a more complex structure near the resonance speed compared with the smoother bi-Maxwellian distribution in panel (d).
The black arrows mark the diffusion directions estimated from VDF.
In the measured VDF shown in panel (c), resonant particles (gray band) diffuse across constant-kinetic-energy contours toward smaller kinetic energies, as indicated by the black arrow.
Such motion corresponds to a transfer of particle energy to the waves and is associated with instability.
By contrast, in the bi-Maxwellian case shown in panel (d), resonant particles diffuse across constant-kinetic-energy contours toward larger kinetic energies, implying that wave energy is transferred to the particles and the mode is damped.

\begin{figure}[!ht]
    \centering
    \includegraphics[width=1.0\linewidth]{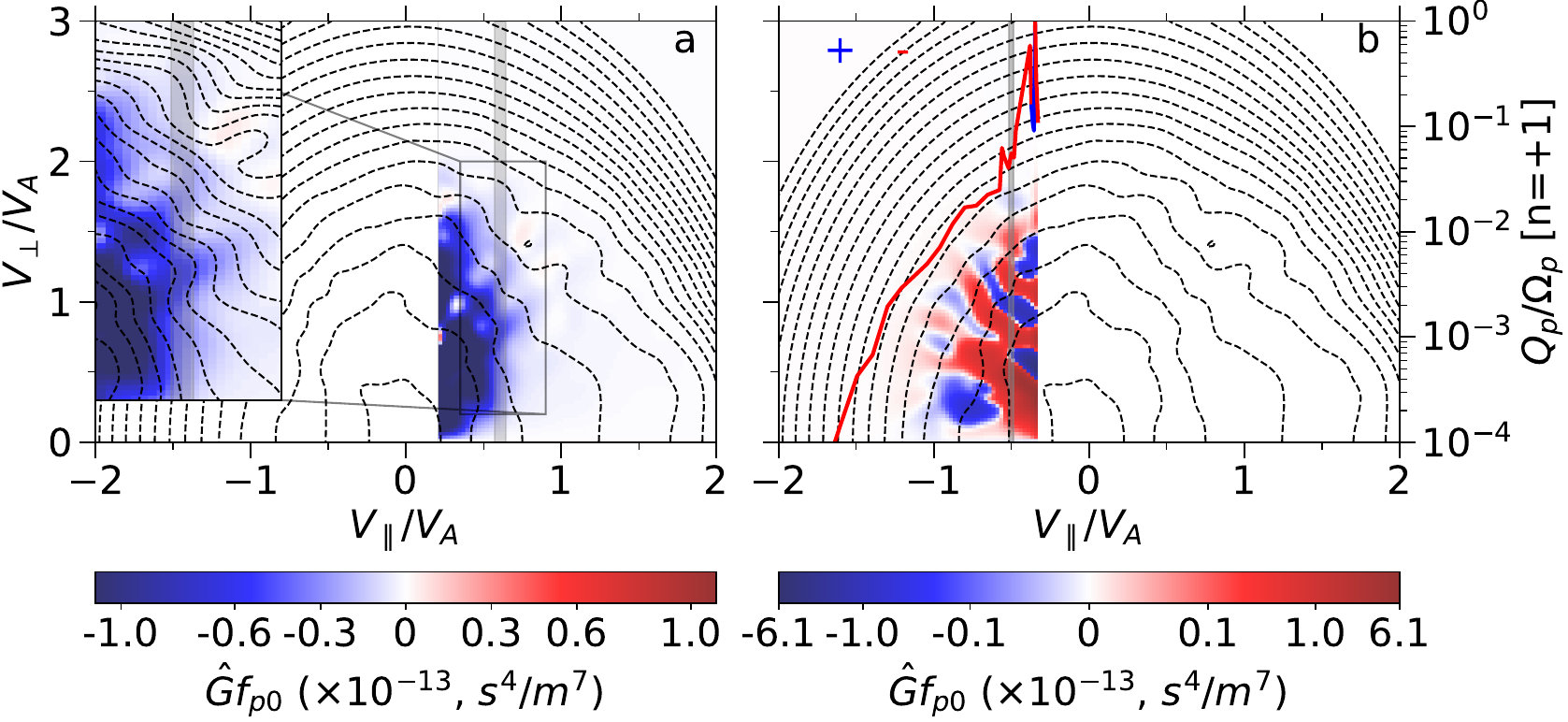}
    \caption{Proton phase-space signatures of resonant wave--particle interactions.
    The dashed contours show the proton VDF in velocity space, same as panel (d) of Figure~\ref{fig:VDF_Construction}.
    The color map indicates the operator $\hat{G} f_{0p}$ for the (a) $n = 0$ and (b) $n = +1$ resonances, with red regions contributing to wave growth (instability) and blue regions to damping.
    In panel (b), the red and blue solid curve traces the proton heating rate $Q_p / \Omega_p$ as a function of $v_\parallel$ for the $n=+1$ resonance, obtained by converting the $k$-dependent heating rate (panel (e) of Figure~\ref{fig:heating_rates}) through Equation~(\ref{eq:resonance}), and is referenced to the right-hand vertical axis.
    }
    \label{fig:G_operator_case1}
\end{figure}

The resonance paths in Figure~\ref{fig:resonance_path_221023} illustrate the diffusive flux of particles in velocity space due to wave--particle interactions but do not show the detailed influence of VDF structures on the plasma response.
To investigate this mechanism in more detail, we evaluate the diffusive operator $\hat{G} f_{j0}$ across velocity space.
Equation~(\ref{eq:growth_rate}) shows that the sign of $\gamma_j$ is the same as the sign of $\hat{G} f_{j0}$, as long as $\mathcal{E}>0$.
Mapping $\hat{G} f_{j0}$ across velocity space provides a representation of the contributions of specific VDF structures to $\gamma_j$.

Figure~\ref{fig:G_operator_case1} shows $\hat{G} f_{p0}$ overplotted on the measured proton VDF for both the $n=0$ (panel (a)) and $n=+1$ (panel (b)) resonances.
Positive values of $\hat{G} f_{p0}$ (red) indicate regions that make a positive contribute to $\gamma_p$ in Equation~\ref{eq:growth_rate}, whereas negative values (blue) indicate a negative contribution to $\gamma_p$.
For the $n=0$ resonance shown in panel (a), the mapping reveals a localized reduction in the damping contribution (indicated by the lighter blue shading) within the resonant $v_\parallel$ range (gray span).
This reduction corresponds to the structural deviations and softened gradients observed in the measured VDF.
Panel (b) shows mapping of the $n=+1$ resonance, where the distribution consists of both blue and red regions.
The blue regions correspond to recessed features of the VDF, while the red regions correspond to protruding features.
The red and blue solid curve traces the proton heating rate for the $n=+1$ resonance as a function of $v_\parallel$ in case (i).
The curve is derived from the heating rate as a function of wavenumber (panel (e) of Figure~\ref{fig:heating_rates}) through the resonance condition in Equation~(\ref{eq:resonance}).
When the curve is negative (red), particles lose energy to the wave, contributing to wave instability; when it is positive, particles gain energy from the wave, leading to wave damping.

\subsection{Coherence Signatures}

\begin{figure}[!ht]
    \centering
    \includegraphics[width=1.0\linewidth]{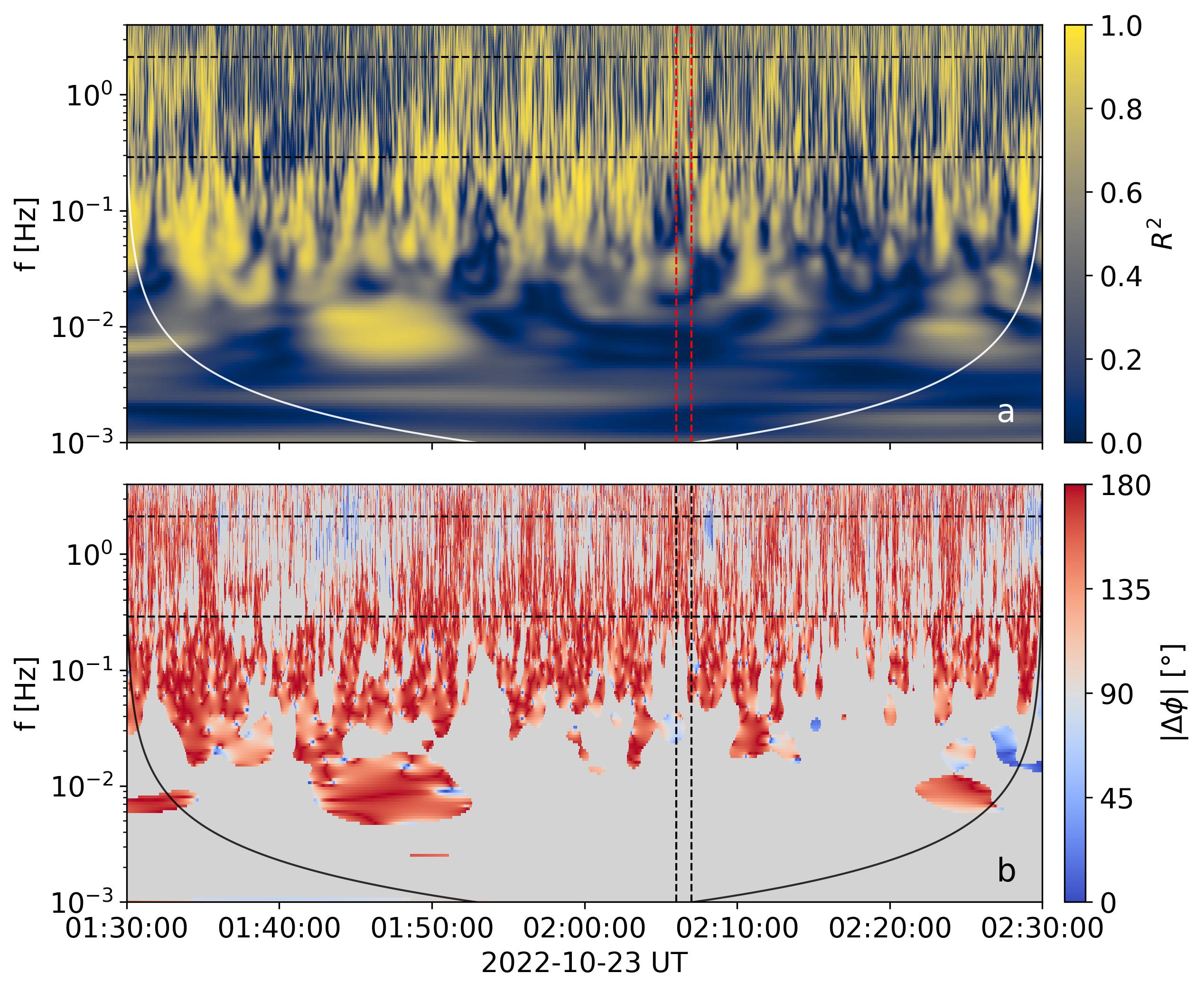}
    \caption{ (a) Squared wavelet coherence $R^2$ (Equation~(\ref{eq:coherence_R2})) between $n_{rpw}$ and $B_\parallel$.
    (b) Absolute cross-wavelet phase angle $|\Delta \phi|$ (Equation~(\ref{eq:coherence_phase})) between the two signals, where a phase shift of $180^\circ$ indicates anti-correlation.
    The solid curves (white in (a), black in (b)) denote the cone of influence in both panels.
    The horizontal black dashed lines mark the spacecraft-frame frequency range for which ALPS predicts the maximum growth of the unstable IA modes (Equation~(\ref{eq:max_growth_f})).
    The vertical dashed lines (red in (a), black in (b)) indicate the time interval over which the VDF is averaged.
    }
    \label{fig:Coherence}
\end{figure}

Figure~\ref{fig:Coherence} shows the results of the wavelet coherence analysis described in Section~\ref{sec: coherence_method}.
Panel (a) displays the squared wavelet coherence $R^2$ (Equation~(\ref{eq:coherence_R2})), where yellow indicates intervals of high coherence between $n_{rpw}$ and $B_\parallel$. 
Panel (b) displays the cross-wavelet phase angle $|\Delta\phi|$ (Equation~(\ref{eq:coherence_phase})), which shows whether this coupling represents a correlation (blue) or an anti-correlation (red) between $n_{rpw}$ and $B_\parallel$.
For comparison with the dispersion relation, we convert the wavenumber at maximum growth, $k_{\gamma, \max}$, obtained from ALPS (i.e., the wavenumber that yields the maximum $\gamma$ in case (i) of Figure~\ref{fig:dispersion_relation_221023}), into the frequency in the spacecraft reference frame as
\begin{equation}
    f \simeq \frac{k_{\gamma, \max} |\bm{{U}}_{sw}| \cos (\theta_{kV})}{2 \pi},
    \label{eq:max_growth_f}
\end{equation}
where $\bm{{U}}_{sw} = (\bm{{U}}_p n_p + \bm{{U}}_\alpha n_\alpha) / (n_p + n_\alpha)$ is the solar wind bulk velocity in the spacecraft frame, and $\theta_{kV}$ is the angle between $\bm{k}$ and $\bm{{U}}_{sw}$.
Since $\bm{k}$ is constrained to a cone of half-angle $\theta_{kB} = 30^{\circ}$ around the magnetic field $\boldsymbol B$ in our ALPS analysis, and the measured angle $\theta_{BV}$ between $\bm{B}$ and $\bm{{U}}_{sw}$ is known from direct measurement, the angle between $\bm{k}$ and $\bm{{U}}_{sw}$ satisfies $|\theta_{BV} - \theta_{kB}| \leq \theta_{kV} \leq |\theta_{BV} + \theta_{kB}|$.
Accordingly, our predictions map to a range in spacecraft-frame frequency that is defined by the maximum and minimum frequencies obtained from Equation~(\ref{eq:max_growth_f}) when evaluated across this $\theta_{kV}$ range. We indicate this range in spacecraft frequency through horizontal black dashed lines in Figure~\ref{fig:Coherence}. 
Our ALPS predictions of unstable waves map to the spacecraft-frame frequency range between these two lines.
The vertical dashed lines (red in panel (a), black in panel (b)) indicate the time interval when the VDF shown in Figure~\ref{fig:VDF_Construction} is recorded.

In Figure~\ref{fig:Coherence} (a), intermittent patches of high $R^2$ occur primarily between $10^{-2}$ and $10^0$~Hz.
From 02:06:00 UT to 02:07:00 UT, the interval to which we apply our ALPS analysis (red dashed lines), $R^2$ is large within the frequency range marked by the black dashed lines.
As shown in panel (b), this specific region of high coherence corresponds to a phase angle of about $180^\circ$ (red), confirming that the density and parallel magnetic-field fluctuations are anti-correlated.

\section{Discussion}
\label{sec: discussion}

We now interpret our results and discuss the limitations of our analysis.

\subsection{Resonance mechanisms regulated by VDF structures}

Independent indicators suggest that the fluctuations studied here correspond to the IA branch.
First, the dispersion relation follows the expected linear trend $\omega_r \simeq k_\parallel c_s$.
Second, the correlation coefficient between fluctuations in the plasma density and in the magnitude of the magnetic field is $-0.698$, consistent with the slow-mode-like polarization associated with IA waves.
Third, the wavelet coherence shows strong anti-correlation between $\delta B_\parallel$ and $\delta n$ in the frequency range at which our ALPS analysis reveals a maximum $\gamma$.

Figure~\ref{fig:dispersion_relation_221023} shows that $\alpha$-particles contribute little to the mode behavior: using the measured $\alpha$-particle distribution produces almost no change to the dispersion relation.
This behavior is reasonable given their relatively low density, weak anisotropy, and sub-Alfv\'enic drift speed in the observed case. 
When the measured proton VDF is used, an instability appears near $k d_p \simeq 1$ even with $T_e \simeq T_p$.
This behavior indicates that fine-scale structures in the proton VDF are responsible for destabilizing the mode.

To understand how proton structures affect the mode behavior, we examine the heating rates and diffusion paths in Sections~\ref{sec: Dispersion_resonance} and \ref{sec: velocity_space_diffusion}.
Figure~\ref{fig:heating_rates} shows that the instability arises from two concurrent effects: a reduction in proton heating through the $n=0$ resonance and the onset of proton cooling through the $n=+1$ resonance compared to the bi-Maxwellian case.
Figure~\ref{fig:resonance_path_221023} illustrates the underlying physics.
For the $n=0$ resonance shown in panel (a), the measured VDF exhibits a localized softening of the velocity gradients (indicated by the red arrow) that overlaps with the resonant parallel velocity range ($v_{res}$, gray band).
This deviation from the steep gradients seen in the bi-Maxwellian case (panel (b)) leads to the suppression of Landau damping.
For the $n=+1$ resonance shown in panel (c), the measured proton VDF exhibits a complex structure near $v_{res}$ (gray band), and the majority of the resonant particles diffuse inward towards smaller kinetic energy,  driving instability.
In the bi-Maxwellian case in panel (d), by contrast, particles diffuse towards greater kinetic energy, which corresponds to damping.

The fine-scale structures in the VDF significantly influence the resonant interaction, as illustrated by the distribution of $\hat{G} f_{p0}$ shown in Figure~\ref{fig:G_operator_case1}.
At a given $v_\parallel$, particles may either lose or gain energy depending on the sign of $\hat{G} f_{p0}$.
For the $n=0$ resonance in panel (a), the regions with lighter blue shading within the resonant velocity range suggest that the softened gradients in the measured VDF lead to a localized reduction in the damping contribution.
For the $n=+1$ resonance in panel (b), protruding structures in the VDF tend to yield positive $\hat{G} f_{p0}$ and contribute to instability, whereas recessed structures correspond to negative $\hat{G} f_{p0}$ and contribute to damping.
The influence of the protruding features is stronger than that of the recessed ones, resulting in a net transfer of energy from particles to the wave.

A key result of our study is the finding that fine-scale structures of the proton VDF near the $n=0$ resonance speed modify the expected Landau and transit-time damping rates of IA waves.
In models based on (bi-)Maxwellian distributions, strong damping of IA waves occurs unless $T_e \gg T_p$. 
These models require large $T_e$ to separate the resonance speed ($\sim c_s$) of IA waves according to Equation~(\ref{eq:sound_speed}) from the thermal speed $w_{\parallel p}$ of the protons, at which many protons are available to resonate, as otherwise, proton damping would strongly damp the wave \citep{barnes1966collisionless, feldman1975solar, gary1993theory}.
However, our results demonstrate that, even when $T_e \simeq T_p$ as commonly observed in the solar wind, the proton VDF can possess local gradients near the $n=0$ resonance speed that suppress the net energy exchange with the wave.
In other words, the plasma remains transparent to compressive waves not because the resonance speed lies outside the proton core, as in the case with $T_e \gg T_p$, but because the fine-scale structure of the VDF reduces the efficiency of resonant damping.
This finding suggests that solar-wind VDFs can dynamically adjust to support the propagation of IA-like modes under near-equal ion and electron temperatures.

\subsection{Possible Origins of the Fine-scale structures}

Alfv\'en/ion-cyclotron (A/IC) waves represent a natural candidate for generating the observed fine-scale structures, as they mainly interact with protons through the $n=+1$ cyclotron resonance \citep{marsch2003ion, ofman2007two, verscharen2019multi}.
The dissipation of A/IC waves can heat protons on their journey from the corona to interplanetary space, producing non-Maxwellian features in the proton VDFs \citep{cranmer2014ensemble}.
Under low-$\beta$ conditions, obliquely propagating A/IC waves efficiently reshape proton distributions through cyclotron resonance, driving perpendicular heating and anisotropy \citep{chandran2010resonant}.
Motivated by these findings, we examine the A/IC wave in our interval and test its stability.
Our hypothesis is that if it were strongly damped, its dissipation could drive the formation of fine-scale structures that subsequently make the IA wave unstable.

\begin{figure}
    \centering
    \includegraphics[width=1.0\linewidth]{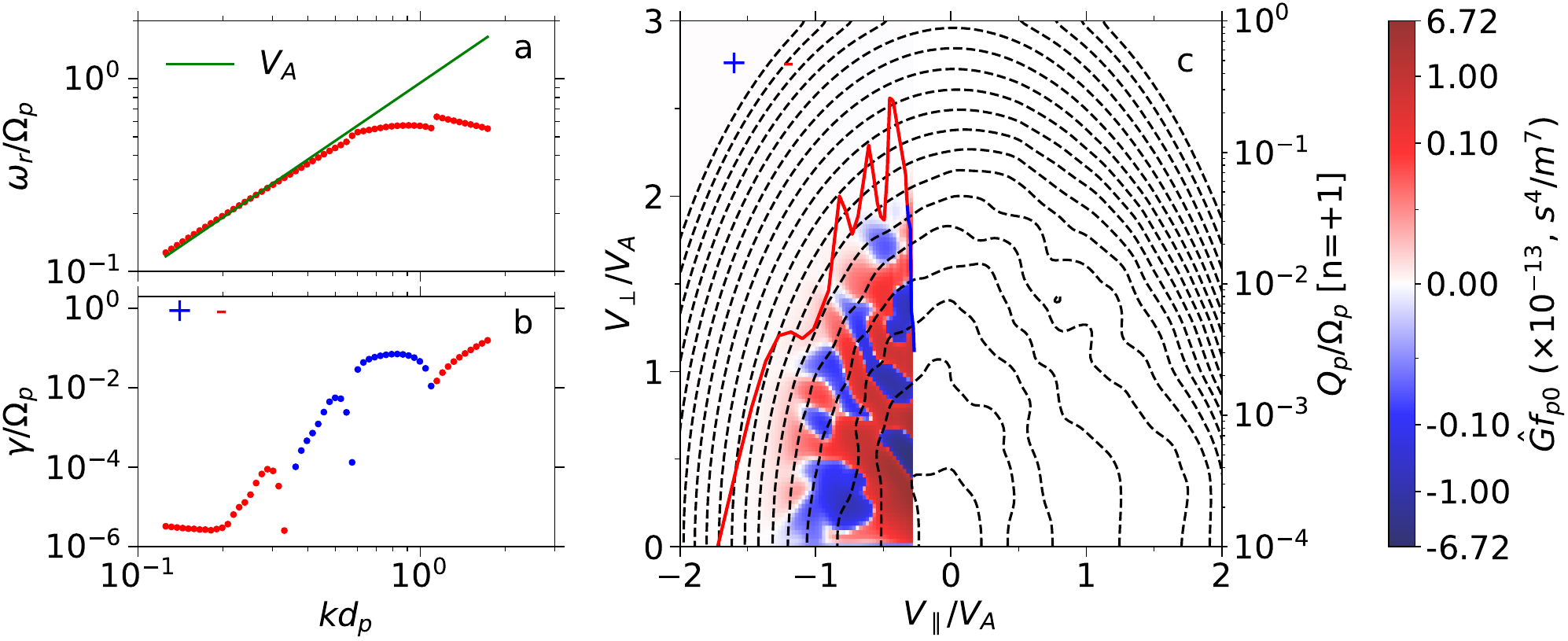}
    \caption{Stability analysis of the A/IC mode for the measured VDF.
    (a) Real frequency $\omega_r / \Omega_p$ as a function of $k d_p$; the green line shows the Alfv\'enic dispersion relation $\omega = k_\parallel V_A$.
    (b) Corresponding growth rate $\gamma / \Omega_p$ versus $k d_p$.
    (c) Diffusive operator $\hat{G} f_{p0}$ (color) of the A/IC mode overlaid on the measured VDF, shown in the same format as Figure~\ref{fig:G_operator_case1}.}
    \label{fig:ICW}
\end{figure}

Figure~\ref{fig:ICW} shows the stability analysis of the A/IC mode for the same measured VDF as in Figure~\ref{fig:G_operator_case1}.
To track the mode, ALPS performs a parallel scan, corresponding to the geometry of parallel-propagating A/IC waves.
Panels (a) and (b) present the dispersion relation, showing that the real frequency follows the expected Alfv\'enic scaling $\omega_r \simeq k_\parallel \mathrm{V_A}$ at small wavenumbers, with widespread positive growth rates indicative of instability.
Panel (c) displays the diffusive operator $\hat{G}f_{p0}$ calculated using the A/IC dispersion relation, overlaid on the measured VDF.
The fine-scale structures that drive the A/IC mode unstable (positive $\hat{G} f_{p0}$ parts in panel (c) of Figure~\ref{fig:ICW}) are located in a velocity-space region that overlaps with the region contributing to the growth of the IA mode.
This suggests that the A/IC instability is triggered by the structures, rather than generating them.

Besides A/IC waves, several other processes may contribute to the formation of the relevant fine-scale structures in our proton VDF.
Turbulence is one such candidate: strong solar wind turbulence produces intermittent coherent structures, such as current sheets and velocity shears, which distort local particle distributions through phase-space mixing and nonlinear interactions \citep{servidio2012local, Servidio2017Cascade, pezzi2018velocity, camporeale2018coherent, larosa2025velocity}.
Another potential driver is stochastic heating, which scatters ions in velocity space, breaks magnetic-moment conservation, and broadens their VDFs \citep{chandran2010perpendicular, vech2017nature, martinovic2020enhancement}.
Further investigation is needed to determine which of these mechanisms, if any, are responsible for the observed fine-scale structures that drive the IA wave unstable.

\subsection{Uncertainties and Limitations of our Analysis}

The GMM separation inherently introduces errors, as it assumes that each particle population can be represented by a Gaussian in velocity space.
While \cite{de2023innovative} assess the performance of GMM in separating different solar-wind populations using synthetic distributions, quantifying its error on real in-situ data remains challenging.
According to their synthetic-data tests, distinguishing the proton beam from the proton core is usually challenging with GMM, whereas the $\alpha$-particle population can be more reliably distinguished.
Since our analysis only requires the combined proton population and the combined $\alpha$-particle population, this limitation has little impact on our results.

Another limitation of our analysis emerges from our construction of the collared VDF.
Unlike parametric approaches that model the proton and $\alpha$-particle populations using prescribed analytic forms across the velocity space, our method prioritizes direct use of the measured data: wherever PAS measurements are available in velocity space, they are treated as the true representation of the VDF. 
This focus on the use of measured data aligns with the approach by \cite{klein2026ion}, who also employ a ``collar" method that uses bi-Maxwellian fits of core and beam components to interpolate and extend the distribution towards high speeds.
Applying our collar method with different smoothing widths of the Gaussian kernel reveals that, although small quantitative differences appear in the resulting dispersion relations, the overall physical behavior remains unchanged.
Although the collared distribution can not fully represent the true solar-wind VDF, it remains consistent with the measurements and provides a physically sensible approximation of the plasma. 
In this sense, we see our combined approach of using measured values whenever available and a collared extension when measured values are not available as an improvement on bi-Maxwellian models that assume a pre-described shape of the VDF throughout all of velocity space.

Our ALPS analysis assumes that the VDFs are gyrotropic, which may not hold in the presence of strong fine-scale structures or near rapidly evolving plasma features.
The gyrotropic assumption is required for quasi-linear stability analysis, but it introduces discrepancies between the kinetic physics of the full three-dimensional VDFs and their reduced two-dimensional gyrotropic representations \cite[e.g.,][]{gedalin2015collisionless, bessho2016electron, lapenta2017origin}.

Instrumental effects may also contribute to uncertainty.
The PAS instrument measures VDFs by sequentially scanning through energy and elevation bins; for each energy step, all elevation bins are sampled before moving to the next energy.
This sweep profile can occasionally cause two neighboring points in gyrotropic velocity space to be sampled at different times, potentially appearing as distinct values and introducing artificial small-scale structure into the measured VDFs.

\section{Conclusions}
\label{sec: conclusions}

This study investigates how fine-scale structures in proton and $\alpha$-particle VDFs regulate the growth and damping rates of IA waves in the solar wind based on high-resolution measurements from Solar Orbiter.
By applying GMM separation to measured VDFs and analyzing the resulting populations with ALPS, we assess the influence of deviations from idealized bi-Maxwellian distributions on wave growth and energy conversion in IA waves.

Our results reveal that the fine structure of the proton VDF near the $n=0$ resonance speed affects the efficiency of Landau and transit-time damping significantly.
Instead of being strongly damped when $T_e \simeq T_p$, as predicted by (bi-)Maxwellian models  \citep{barnes1966collisionless, feldman1975solar, gary1993theory}, IA waves propagate without strong damping when the VDFs exhibit local gradients that suppress resonant damping, even when the proton and electron temperatures are nearly equal.

The fine-scale structure of the proton VDF at $v_\parallel < 0$ can locally invert the velocity space gradient near the resonance speed associated with $n=+1$, providing a source of free energy that can contribute to IA wave growth.
Our results demonstrate that the measured fine-scale structure in the proton VDF influences the wave in two ways, suppressing damping at the $n=0$ resonance and enabling growth at the $n=+1$ resonance.
This finding highlights the crucial role of fine-scale velocity-space structures in regulating the stability and propagation of compressive fluctuations in the solar wind.

Other wave modes and nonlinear processes may contribute to the shaping of the proton VDF in ways that make the plasma either transparent or unstable to compressive waves.
In particular, the A/IC mode in this interval is unstable, suggesting that the observed fine-scale structures in the VDF are not a consequence of its dissipation, but they may form through kinetic wave--particle interactions that involve the unstable A/IC waves.
Further work is needed to determine which additional processes, such as intermittent turbulence or stochastic heating, govern the formation and evolution of the relevant structures in the proton VDF.

Our work demonstrates that fine-scale structure in the solar-wind proton VDF has a decisive influence on the damping and growth of compressive plasma waves.
Our combined GMM-ALPS approach provides a powerful framework for the evaluation of the kinetic impacts of measured velocity-space features.
Our approach opens new avenues for the identification of the microphysical processes that regulate energy transfer and wave--particle interactions in the heliosphere.
Future applications at different heliocentric distances and in different solar-wind regimes may further clarify how fine-scale structures in the VDFs of the particles shape the evolution of space plasmas throughout the heliosphere.

\section*{Acknowledgments}
H.R. and X.W. are supported by STFC grant ST/Y509784/1. 
D.V. and C.J.O. are supported by STFC grant ST/W001004/1 and UKRI grant UKRI1204. 
C.I. is supported by STFC grant ST/X508858/1. 
K.G.K. is supported in part by NASA grant 80NSSC24K0724.
This research was supported by the International Space Science Institute (ISSI) in Bern, through ISSI International Team project \#612 (Excitation and Dissipation of Kinetic-Scale Fluctuations in Space Plasmas) led by K.~G.~Klein. 
The authors acknowledge helpful discussions with Rossana De Marco.
Solar Orbiter is a mission of international cooperation between ESA and NASA, operated by ESA. Solar Orbiter SWA data were derived from scientific sensors that were designed and created and are operated under funding provided by numerous contracts from UKSA, STFC, the Italian Space Agency, CNES, the French National Centre for Scientific Research, the Czech contribution to the ESA PRODEX program, and NASA.  Solar Orbiter SWA operations at the UCL/Mullard Space Science Laboratory are currently funded by UKRI grant UKRI919.
The codes used in this study \citep{Ran:2026} are available under an open-source MIT License at \url{https://github.com/RanHao1999/SWA-Data-Analysis}, with the specific version used for this work archived as \dataset[SWA-Data-Analysis v1.0.0]{https://doi.org/10.5281/zenodo.18902395}.
The ALPS project received support from UCL's Advanced Research Computing Centre through the Open Source Software Sustainability Funding, and the code \citep{ALPS:2023} is available via an open source BSD 2-Clause License at \url{https://github.com/danielver02/ALPS} with a full tutorial on its use at \url{https://danielver02.github.io/ALPS/}.

\bibliography{sample701}{}
\bibliographystyle{aasjournalv7}



\end{CJK*}
\end{document}